\documentstyle [12pt,epsf]{article}

\input{epsf}
\begin{document}
\begin{titlepage}
\begin{center}

 \vspace{-0.1in}

{\large \bf Functional Methods in \\
the Generalized Dicke Model}\\
 \vspace{.3in}{\large\em M.~Aparicio Alcalde,\,\footnotemark[1]
A.~L.~L.~de~Lemos\,\footnotemark[2]
 and N.~F.~Svaiter
\footnotemark[3]}\\
\vspace{.2in}
 Centro Brasileiro de Pesquisas F\'{\i}sicas,\\
 Rua Dr. Xavier Sigaud 150,\\
 22290-180, Rio de Janeiro, RJ Brazil \\

\subsection*{\\Abstract}
\end{center}

\baselineskip .16in

The Dicke model describes an ensemble of $N$ identical two-level
atoms (qubits) coupled to a single quantized mode of a bosonic
field. The fermion Dicke model should be obtained by changing the
atomic pseudo-spin operators by a linear combination of Fermi
operators. The generalized fermion Dicke model is defined
introducing different coupling constants between the single mode
of the bosonic field and the reservoir, $g_{1}$ and $g_{2}$ for
rotating and counter-rotating terms respectively. In the limit $N
\rightarrow \infty$, the thermodynamic  of the fermion Dicke model
can be analyzed using the path integral approach with functional
method. The system exhibits a second order phase transition from
normal to superradiance at some critical temperature with the
presence of a condensate. We evaluate the critical transition
temperature and present the spectrum of the collective bosonic
excitations for the general case $(g_{1}\neq 0$ and $g_{2}\neq
0)$. There is quantum critical behavior when the coupling
constants $g_{1}$ and $g_{2}$ satisfy
$g_{1}+g_{2}=(\omega_{0}\,\Omega)^{\frac{1}{2}}$, where
$\omega_{0}$ is the frequency of the mode of the field and
$\Omega$ is the energy gap between energy eigenstates of the
qubits. Two particular situations are analyzed. First, we present
the spectrum of the collective bosonic excitations, in the case
$g_{1}\neq 0$ and $g_{2}=0$, recovering the well known results.
Second, the case $g_{1}=0$ and $g_{2}\neq 0$ is studied. In this
last case, it is possible to have a superradiant phase when only
virtual processes are introduced in the interaction Hamiltonian.
Here also appears a quantum phase transition at the critical
coupling $g_{2}=(\omega_{0}\,\Omega)^{\frac{1}{2}}$, and for
larger values for the critical coupling, the system enter in this
superradiant phase
with a Goldstone mode. \\
\vspace{0,1in}
PACS numbers: 42.50.Fx, 05.30.Jp

\footnotetext[1]{e-mail: \,aparicio@cbpf.br}
\footnotetext[2]{e-mail:\,\,aluis@cbpf.br}
\footnotetext[3]{e-mail:\,\,nfuxsvai@cbpf.br}

\end{titlepage}
\newpage\baselineskip .18in

\section{Introduction}

\quad $\,\,$The aim of the present paper is to investigate the
thermodynamic in a generalization of the Dicke model \cite{dicke},
using the path integral approach with the functional integration
method. We are considering the question of how does the
counter-rotating terms of the interaction Hamiltonian contribute
to the system exhibits a phase transition from normal to
superradiance at some critical temperature $\beta^{-1}_{c}$.

We are particularly interested in consider $N$ identical two-level
atoms interacting with a quantized bosonic field, i.e., the
spin-boson problem. The most discussed situation in the literature
is a system of atoms interacting with a reservoir which consists
of a surrounding quantized electromagnetic field. In this
situation the frequencies associated with the electromagnetic
field are continuously distributed and the atomic systems are
characterized by a denumerable set of frequencies. Nevertheless,
different situations has been extensively studied. One is the case
of electromagnetic field in cavities where the field modes are
described by a set of denumerable frequencies \cite{meshede}
\cite{meystre}. One particular example is the Jaynes-Cummings
model \cite{ja1}, where was investigated the dynamics of a single
two-level atom interacting with one mode of a quantized bosonic
field in an ideal lossless cavity. The eigenstates of the model,
i.e., the dressed states of the two-level atom-bosonic mode were
first obtained by these authors. After these important papers, the
spin-boson problem has been studied by many authors, receiving
considerable discussions. The development of new techniques has
lead to increase the interest in spontaneous radiations effects in
such systems. More recently, this kind of model has been used also
to analyze dissipation in quantum computers. A model describing a
two-level atom coupled to a reservoir of harmonic oscillators was
introduced by Di Vicenzo \cite{di}, to study decoherence in
quantum computers. This model presents destruction of quantum
coherence without decay of populations, owing to a quite
particular coupling between the qubit and the reservoir.

In the following we are discussing the Dicke model, where a single
quantized mode of a bosonic field interacts with a reservoir of
$N$ identical two-level atoms (qubits) at temperature
$\beta^{-1}$. One method to obtain information from the small
system of interest, eliminating the reservoir degrees of freedom
is to use the master equations, introducing the reduced density
operator \cite{dekker} \cite{livro}. An alternative and
non-perturbative powerful tool to deal with the elimination of the
reservoir degrees of freedom, is the path integral approach with
functional integration method \cite{pop2}. In a general situation,
this method allow us to obtain the semiclassical equation of
motion and also gives a systematic way for the evaluation of
quantum corrections.

Although functional integrals have been used in different area of
physics, as for example quantum field theory and also statistical
mechanics, in the past this technique have no found much use in
quantum optics. An exception is the Zardecki paper \cite{zurdeki}.
Other paper that deserves to call attention is the Hillery and
Zubaire \cite{hillery1} work. In this paper the path integral
approach in quantum optics was developed, where systems with a few
bosonic modes were studied. Using the
Schrodinger-Glauber-Sudarshan's coherent states \cite{schrodinger}
\cite{glauber1} \cite{glauber2} \cite{glauber3} \cite{sudarshan}
\cite{ja2} representation of these modes, these authors studied
the propagator in a single mode system and also the case of $N$
modes of a bosonic field. Since, in the Dicke model, the coupling
between the $N$ qubits and the single quantized mode of the
bosonic field is linear, in order to analyze the nonanalytic
behavior of thermodynamic quantities, the integration over the
degrees of freedom of the environment can be performed exactly.

To apply the path integral approach with  functional method, as a
technique to investigate the thermodynamic of the Dicke model, two
steps are mandatory. First, it is necessary to change the atomic
pseudo-spin operators of the model by a linear combination of
Fermi operators to define the fermion Dicke model \cite{kondo}
\cite{chiang}. A generalization of this model can be achieved
introducing the counter-rotating terms in the interaction
Hamiltonian. Second, the thermodynamic limit ($N\rightarrow
\infty$), where $N$ is the number of qubits must be taken
\cite{pri} \cite{popov}. Therefore, studying the fermion Dicke
model with counter-rotating terms, describing virtual processes,
we are interested to evaluate the critical transition temperature
of the model. In this generalized model, one may ask the
importance of the contribution of each of these processes, real
ones and virtual ones in the formation of the condensate.

A point that it is worth commenting upon is the relationship
between our questions concerning the contributions of the real and
virtual processes in the superradiant phase transition and how
does the vacuum fluctuations and the self-reaction contribute to
the radiative processes in atoms. Many authors \cite{knight1}
\cite{milonni1}  \cite{milonni2} used the Heisenberg picture to
identify the vacuum and the self-reaction contributions in
radiative processes in atoms. Although there was some attempts to
separate the contribution of the vacuum fluctuations and radiative
reaction in the spontaneous decay \cite {fain} \cite{eberly},
Dalibard et al \cite {roc1} \cite{roc2} discussed how does the
magnitude of these separate effects can varied by means of the
particular ordering chosen for atomic and creation and
annihilation field operators. When normal ordering is used, the
source field gives the total contribution to the decay, and for
anti-normal ordering, both the source field and vacuum field gives
equal contributions to the spontaneous decay. Finally, if the
symmetric ordering is chosen, the vacuum and the source field
gives different contributions for the decay. There is no ordering
where the spontaneous emission can be entirely associated to the
vacuum field \cite{mil}.

As we already emphasized, we are interested to study the
nonanalytic behavior of thermodynamics quantities near a phase
transition in the fermion Dicke model where the counter-rotating
terms are also presented. Introducing different coupling
constants, $g_{1}$ and $g_{2}$ for rotating and counter-rotating
terms respectively, we are able to identify the contribution of
each of the processes, real ones and virtual ones in the formation
of the condensate. The order parameter of the transition is the
expectation value of the number of excitations associated to the
mode of the bosonic field per atom. We evaluate the critical
transition temperature of the model and present the spectrum of
the colective bosonic excitations in the general case, for the
case $g_{1}\neq 0$ and $g_{2}=0$ and also $g_{1}=0$ and $g_{2}\neq
0$. In both cases it appear a critical behavior with Godstone
(gapless) modes. In the last case, it appears a quantum phase
transition \cite{sachdev} at critical coupling
$g_{2}=(\omega_{0}\,\Omega)^{\frac{1}{2}}$. Therefore, in the
fermion Dicke model with the counter-rotating terms it is possible
to have a quantum fase transition and a superradiant phase, where
the radiation rate on the number $N$ of atoms becomes a quadratic
dependence \cite{yarunin}, in opposite to the situation of $N$
atoms radiating incoherently, where the radiation rate is
proportional on the number of the atoms $N$.

At this point we would like to make a summary of results
concerning the thermodynamic of the Dicke model. An important
result were obtained by Hepp and Lieb \cite{hepp}. These author
presented the free energy of the model in the thermodynamic limit.
For a sufficiently large value for the coupling constant between
the qubits and the single quantized mode of the bosonic field, the
model present a second order phase transition from the normal to
the superradiant phase. Latter, using a coherent state
representation they generalize some results for atoms of more than
two-levels without assume the rotating-wave approximation. They
also investigated the stability of the model with an infinite
number of bosonic modes \cite{hepp2}. The study of the phase
transitions in the Dicke model was presented also by Wang and Hioe
\cite{wang}, where some of the results obtained by Hepp and Lieb
were rederived. The generalized Dicke model, where the
counter-rotating terms are also present in the interaction
Hamiltonian, was investigated by Hioe \cite{hioe} and also Duncan
\cite{duncan}. Hioe studied the thermodynamic of the generalized
Dicke model with two different coupling constants using the
Schrodinger-Glauber-Sudarshan's coherent states. Duncan proved
that the critical temperature of the generalized model is higher
than the temperature of the original model. Also a bosonization
procedure was employed to study the phase transitions in the
generalized Dicke model. Employing a Holstein-Primakoff mapping
\cite{prim} \cite{papanicolau} \cite{hillery}, which express the
angular momentum in terms of a single bosonic mode, Emary and
Brandes \cite{emary} \cite{emary2} were able to express the
generalized Dicke model in terms of two mode bosonic field. These
authors discussed the relation between the quantum phase
transition and the chaotic behavior that appear in the model for
finite $N$. An indication of quantum chaos is the change in energy
level-spacing statistics from Poissonian to being described by the
Gaussian ensemble of random matrix theory. This chaotic behavior
was discussed also in the Jaynes-Cummings model by Graham and
Hoherbach \cite{gra} \cite{gra2} and Lewenkopf et al.
\cite{nemes}, in the situation where the counter-rotating terms
are present in the interaction Hamiltonian \cite{muller}
\cite{finney}, since the seminal paper of Milonni et al.
\cite{seminal}

We would like to stress that we are not doing any distinction
between the elementary unit of quantum information, the qubit and
the two-level atom. Our intention in this paper is to review some
important spin-boson models and also the basic ideas of the path
integral approach with functional methods that can be used in
these systems. This paper is organized as follows. In section II
we discuss two-level atoms-boson field interaction Hamiltonians.
In section III the Hilbert space for the fermion Dicke model is
discussed. In the section IV the path integral with functional
integral method is applied to study the thermodynamic of the
generalized fermion Dicke model. We evaluate the critical
transition temperature of the model for the general case
$(g_{1}\neq 0$ and $g_{2}\neq 0)$ and also present the spectrum of
the collective bosonic excitations, for the case $g_{1}\neq 0$ and
$g_{2}=0$ and also for the case $g_{1}=0$ and $g_{2}\neq 0$.
Conclusions are given in section V. In the paper we use
$k_{B}=c=\hbar=1$.

\newpage
\section{The two-level atoms-Bose field Hamiltonians}\

In order to describe the dynamics of the reservoir and the small
system, we have to introduce the Hamiltonian governing the
interaction of the quantized bosonic field with free non-identical
two-level atoms. Free means that there is no interaction between
the two-level atoms (qubits). We are assuming that the qubits are
enclosed in a very large lossless cavity, and the distance between
the qubits is large enough, so that the interaction between the
qubits can be neglected.

The main purpose of this section is to review a number of relevant
qubit-Bose field Hamiltonians. This discussion is standard in the
literature. For a recent treatment see for example the Ref.
\cite{ricardo}. It is worth mentioning that there are a variety of
theoretical models of reservoirs. We have chosen the models that
seems to us most significant and interesting. The first situation
is when the system $S$ is coupled to an infinite number of
harmonic oscillators. In this situation there are two kinds of
reservoir of common interest. The first one is a thermal
reservoir, where we assume that the harmonic oscillators are in
thermal equilibrium at temperature $\beta^{-1}$. The second one is
a squeezed reservoir. The specific system-reservoir model which is
appropriate for the study of several interesting situations is
when the harmonic oscillator bath is constituted by a bosonic
field in free space also in the presence of macroscopic
structures.

Therefore let us consider a bosonic quantum system $S$, with
Hilbert space ${\cal H}^{(S)}$ which is coupled with the reservoir
of qubits, with Hilbert space  ${\cal H}^{(B)}$. Let us assume
that the reservoir is in thermal equilibrium at temperature
$\beta^{-1}$. The bosonic quantum system is a sub-system of the
total system living in the tensor product space ${\cal
H}^{(S)}\,\otimes\,{\cal H}^{(B)}$.

Let us denote by $H_{S}$ the Hamiltonian of the quantized bosonic
field (where the "S" means small system), by $H_{B}$  the free
Hamiltonian of the $N$-qubits (where the "B" means bath) and
$H_{I}$ the Hamiltonian describing the interaction between the
quantized bosonic field and the qubits reservoir. The Hamiltonian
for the total system can be written as
\begin{equation}
H=H_{S}\,\otimes\,I_{B}+I_{S}\,\otimes\,H_{B}+H_{I}\, ,
\label{13}
\end{equation}
where $I_{S}$ and $I_{B}$ denotes the identities in the Hilbert
spaces of the quantized bosonic field and the qubit reservoir.

The free $j-th$ qubit Hamiltonian will be denoted by
$H_{D}^{(j)}$. Therefore we have
\begin{equation}
H_{D}^{(j)}|\,i\,\rangle_{j}=\omega_{i}^{(j)}|\,i\,\rangle_{j}\, ,
\label{14}
\end{equation}
where $|\,i\,\rangle_{j}$ are orthogonal energy eigenstates
accessible to the $j-th$ qubit and $\omega_{i}^{(j)}$ are the
respective eigenfrequencies. Using Eq. (\ref{14}) and the
orthonormality of the energy eigenstates we can write the  $j-th$
qubit Hamiltonian $H_{D}^{(j)}$ as
\begin{equation}
H_{D}^{(j)}=\sum_{i=1}^{2}\,\omega_{i}^{(j)}\biggl(|\,i\,\rangle\,
\langle\, i|\biggr)_{j}\, .
\label{15}
\end{equation}
Throughout the article, for the sake of simplicity in the
notation, we are assuming that the operator $
\biggl(|\,i\,\rangle\, \langle\, i|\biggr)_{j}\,\equiv
|\,i\,\rangle_{j}\,_{j} \langle\, i|,$ and the same convention is
used for the other operators.

Let us define the pseudo-spin operators for each qubit
$\sigma_{(j)}^{\,z}$, and the raising and lowering pseudo-spin
operators $\sigma_{(j)}^{+}$ and $\sigma_{(j)}^{-}$, where each of
these operators are given respectively by
\begin{equation}
\sigma_{(j)}^{\,z}=\biggl(\,|\,2\,\rangle\,\langle\,2|-
\,|1\,\rangle\,\langle\,1|\biggr)_{j}\, , \label{16}
\end{equation}
\begin{equation}
\sigma_{(j)}^{+}=\biggl(\,|\,2\,\rangle\,\langle\,1|\biggr)_{j}
\label{17}
\end{equation}
and finally
\begin{equation}
\sigma_{(j)}^{-}=\biggl(\,|1\,\rangle\,\langle\,2|\biggr)_{j}\, .
\label{18}
\end{equation}
This representation is a second quantization of the qubits.
Combining Eq. (\ref{15}) and Eq. (\ref{16}), the $j-th$ qubit
Hamiltonian can be written as
\begin{equation}
H_{D}^{(j)}=\frac{\Omega^{(j)}}{2}\,\sigma_{(j)}^z+\frac{1}{2}
\biggl(\omega_{1}^{(j)}+\omega_{2}^{(j)}\biggr)\, , \label{19}
\end{equation}
where the energy gap between the energy eigenstates of the $j-th$ qubit
is given by
\begin{equation}
\Omega^{(j)}=\omega_{2}^{(j)}-\omega_{1}^{(j)}\, .
\label{20}
\end{equation}
Shifting the zero of energy to
$\frac{1}{2}(\omega_{1}^{(j)}+\omega_{2}^{(j)})$ for each qubit,
the $j-th$ qubit Hamiltonian given by Eq. (\ref{19}) can be
rewritten as
\begin{equation}
H_{D}^{(j)}=\frac{\Omega^{(j)}}{2}\,\sigma_{(j)}^z\, .
\label{21}
\end{equation}
Note that the pseudo-spin operators $\sigma_{(j)}^{+}$,
$\sigma_{(j)}^{-}$ and $\sigma_{(j)}^z$ satisfy the standard
angular momentum commutation relations corresponding to spin
$\frac{1}{2}$ operators, i.e.,
\begin{equation}
\left[\,\sigma_{(j)}^{+},\sigma_{(j)}^{-}\,\right]=\,\sigma_{(j)}^z\,
, \label{22}
\end{equation}
\begin{equation}
\left[\,\sigma_{(j)}^z,\sigma_{(j)}^{+}\,\right]=2\,\sigma_{(j)}^{+}\,
, \label{23}
\end{equation}
and finally
\begin{equation}
\left[\,\sigma_{(j)}^z,\sigma_{(j)}^{-}\,\right]=-2\,\sigma_{(j)}^{-}\,
. \label{24}
\end{equation}
To introduce the coupling between the bosonic field and the
qubits, let us assume a single quantized mode of a bosonic field
with a linear coupling with the qubits. Therefore, the Hamiltonian
of the $j-th$ qubit $H_{D}^{(j)}$, with the contribution of the
single quantized mode of the bosonic field $H_{S}$, and the
interaction Hamiltonian $H_{I}^{(j)}$ is given by
\begin{eqnarray}
&& I_{S}\,\otimes\,H_{D}^{(j)}+H_{S}\,\otimes\,I_{B}+H_{I}^{(j)}=\nonumber\\
&&
I_S\,\otimes\,\frac{\Omega^{(j)}}{2}\,\sigma_{(j)}^z+\omega_{0}\,b^{\dagger}\,b\,\otimes\,
I_B+g\,\Bigl(b\, +\,
b^{\dagger}\Bigr)\otimes\,\Bigl(\sigma_{(j)}^{+}+\sigma_{(j)}^{-}\Bigr)\,
,
\label{25}
\end{eqnarray}
where the second term in the Eq. (\ref{25}) has the contribution
from the single quantized mode of the bosonic field Hamiltonian
and the last term is the interaction Hamiltonian of the $j-th$
qubit with the one-mode quantized field. In the above equation $g$
is a small coupling constant between the qubit and the single mode
of the bosonic field. The $b$ and $b^{\dagger}$ are the boson
annihilation and creation operators of mode excitations that
satisfy the usual commutation relation rules. The generalization
to $N$ qubits is described by
\begin{eqnarray}
&&I_{S}\,\otimes\,\sum_{j=1}^{N}\,H_{D}^{(j)}+H_{S}\,\otimes\,I_{B}+
\sum_{j=1}^{N}\,H_{I}^{(j)}=\nonumber\\
&& I_S\,\otimes\,\sum_{j=1}^{N}\,
\frac{\Omega^{(j)}}{2}\,\sigma_{(j)}^z+\omega_{0}\,b^{\dagger}\,b\,\otimes\,I_B+
\,\frac{g}{\sqrt{N}} \sum_{j=1}^{N}\,
\Bigl(b+b^{\dagger}\Bigr)\otimes
\Bigl(\sigma_{(j)}^{+}+\sigma_{(j)}^{-}\Bigr)\, . \label{26}
\end{eqnarray}
This Hamiltonian is a possible generalization of the Dicke model.
The interaction Hamiltonian given by Eq.(\ref{26}) is simplified
if we ignore the counter-rotating terms. Considering this
approximation we have

\begin{eqnarray}
&& I_{S}\,\otimes\,\sum_{j=1}^{N}\,
H_{D}^{(j)}+H_{S}\,\otimes\,I_{B}+\sum_{j=1}^{N}\,H_{I}^{(j)}=\nonumber\\
&&
I_S\,\otimes\,\sum_{j=1}^{N}\,\frac{\Omega^{(j)}}{2}\,\sigma_{(j)}^z+
\omega_{0}\,b^{\dagger}\,b\,\otimes\,I_B+
\frac{g}{\sqrt{N}}\sum_{j=1}^{N}\, \Bigl(b\,\otimes\,
\sigma_{(j)}^{+}\,+b^{\dagger}\,\otimes\,\sigma_{(j)}^{-}\Bigr)\,
. \label{28}
\end{eqnarray}
This Hamiltonian describes the original Dicke model in the
situation $\Omega^{(j)}=\Omega, \forall j$. This approximation is
known as the rotating-wave approximation. In the rotating-wave
approximation we ignore energy non-conserving terms in which the
emission (absorption) of a quantum of a quantized field is
accompanied by the transition of one qubit from its lower (upper)
to its upper (lower) state. The rotating-wave approximation
ignores terms in which the $j-th$ qubit raising (lowering)
operators multiplies the mode of the field creation (annihilation)
operator. Although the rotating-wave approximation has been used
to describes the interaction of radiation with matter, as we
stressed, there are many situations where we can not assume the
above mentioned approximation, as for example if we go beyond the
dipole approximation, to describe Casimir-Polder forces
\cite{casimir}, the Lamb-shift and also when atomic systems
generate high intensity fields.

Situations where we can not assume the rotating-wave approximation
has been also extensively discussed by one of the authors. First
Svaiter and Svaiter \cite{ss} \cite{ss2} evaluated the transition
rates of a two-level atom in different kinematic situations
assuming a weak coupling between this small system and a real
massless scalar field. These authors studied the Unruh-Davies
effect \cite{davies} \cite{unruh}, where an accelerated two-level
atom measures a thermal spectrum even if the field is prepared in
the vacuum state. The origin of this effect is the presence of an
event horizon that change virtual processes in real ones. In the
Ref. \cite{fs} still studying the same two-level system coupled to
a real scalar field, Ford et al assumed the presence of two
infinite perfectly reflecting plates which change the vacuum
fluctuations associated to the quantized bosonic field. The image
method and the imaginary time formalism \cite{kubo} \cite{martin}
\cite{fulling}, was used to study radiative processes at finite
temperature. In the Refs. \cite{vitorio} \cite{paola} one of the
authors continue to investigate radiative precesses associated to
the Unruh-Dewitt detector \cite{Dewitt}, in interaction with a
massless scalar field. Being more precise, in the Ref.
\cite{paola}, it was presented the detector's excitation rate when
it is uniformly rotating around some fixed point, when the scalar
field is prepared in the Minkowski vacuum.

Another model, where the behavior is quite interesting from the
mathematical and physical point of view is the one where the
coupling between the $N$ qubits and the bosonic field in a
lossless cavity is intensity dependent. We have
\begin{eqnarray}
&& I_{S}\,\otimes\,\sum_{j=1}^{N}\,
H_{D}^{(j)}+H_{S}\,\otimes\,I_{B}+\sum_{j=1}^{N}\,H_{I}^{(j)}=\nonumber\\
&&
I_S\,\otimes\,\sum_{j=1}^{N}\,\frac{\Omega^{(j)}}{2}\,\sigma_{(j)}^z+
\omega_{0}\,b^{\dagger}\,b\,\otimes\,I_B+
\frac{g}{\sqrt{N}}\sum_{j=1}^{N}\,
\Bigl(b\,(b^{\dagger}\,b)^{\frac{1}{\,2}}\otimes\,
\sigma_{(j)}^{+}\,+b^{\dagger}\, (b^{\dagger}\,b)^{\frac{1}{\,2}}
\otimes\,\sigma_{(j)}^{-}\Bigr)\, . \label{a28}
\end{eqnarray}

Going back to the case of only one qubit, the two-level atom
coupled to a single mode quantized electromagnetic field is known
as the Jaynes-Cummings model. The Jaynes-Cummings model for one
qubit, can be written a as
\begin{eqnarray}
&&
I_{S}\,\otimes\,H_{D}^{(j)}+H_{S}\,\otimes\,I_{B}+H_{I}^{(j)}=\nonumber\\
&&
I_S\,\otimes\,\frac{\Omega^{(j)}}{2}\,\sigma_{(j)}^z+\omega_{0}\,b^{\dagger}\,b\,\otimes\,I_B+
g\Bigl(b\,\otimes\,\sigma_{(j)}^{+}+b^{\dagger}\,\otimes\sigma_{(j)}^{-}\Bigr)\,
.
\label{27}
\end{eqnarray}
Practical realization of this model in the laboratory is not a
problem. First we consider only one two-level atom in the presence
of a quantized bosonic field in a cavity. Assuming that the
frequency $\omega_{0}$ of one of the cavity modes is near-resonant
with the energy gap $\Omega$ of the two-level atom, such situation
generates the following physical model. The two-level atom
effectively interacts only with that mode, and all the other
bosonic modes do not couple with the two-level atom.

This model provides a useful means of studying nonclassical
effects in the interaction between fields and matter, as for
example the phenomenon of collapses and revivals of the Rabi
oscillations in a quantized field that is not in a pure number
state \cite{jc2} \cite{jc3} \cite{jc4} \cite{yoo}. It is important
to point out that the behavior of the two-level atom depends if
the single field mode is quantized or not. In the case of a
classical field the model presents atomic population inversion
with monotonic periodic oscillations. Assuming that the single
field mode is quantized the system presents two different
dynamical behavior. If we prepare the atom in the ground state and
the field mode in a pure number state, it can be shown that the
expectation value of the pseudo-spin operator $\sigma^{z}$ is the
same as in the case of a classical field. If we prepare the atom
in the excited state and the field mode in a pure number state,
the atom periodically returns to the upper state with a definite
Rabi frequency. For the particular case of the atom in an empty
cavity (zero occupation number i.e., $n=0$), the two-level system
undergoes periodic, reversible spontaneous decay. Finally,
preparing the field mode in a coherent state, the dynamics of the
two-level system present quantum collapse and also revival
effects.

It is possible to generalize the Jaynes-Cummings model, still
using the rotating-wave approximation in the following way
\begin{eqnarray}
&&
I_{S}\,\otimes\,H_{D}^{(j)}+H_{S}\,\otimes\,I_{B}+H_{I}^{(j)}=\nonumber\\
&&
I_S\,\otimes\,\frac{\Omega^{(j)}}{2}\,\sigma_{(j)}^z+\omega_{0}\,b^{\dagger}\,b\,\otimes\,I_B+
g\Bigl(b^{2}\otimes\,\sigma_{(j)}^{+}+(b^{\dagger})^{2}
\otimes\sigma_{(j)}^{-}\Bigr)\, . \label{b27}
\end{eqnarray}
In the literature the above model is known as the two-photon
Jaynes-Cummings model \cite{puri} \cite{toor}.

The same idea that has been used in the Eq. (\ref{a28}) can also
be implemented in the Jaynes-Cummings model. The Hamiltonian for
the Jaynes-Cummings model with a intensity-dependent coupling
constant where we are still using the rotating-wave approximation
is given by \cite{jj1} \cite{jj2} \cite{jj3}

\begin{eqnarray}
&&
I_{S}\,\otimes\,H_{D}^{(j)}+H_{S}\,\otimes\,I_{B}+H_{I}^{(j)}=\nonumber\\
&&
I_S\,\otimes\,\frac{\Omega^{(j)}}{2}\,\sigma_{(j)}^z+\omega_{0}\,b^{\dagger}\,b\,\otimes\,I_B+
g\Bigl(b\,(b^{\dagger}b)^{\frac{1}{\,2}}\,\otimes\,
\sigma_{(j)}^{+}+b^{\dagger}\,(b^{\dagger}b)^{\frac{1}{\,2}}\,\otimes\sigma_{(j)}^{-}\Bigr)\,
. \label{t27}
\end{eqnarray}

So far we have discussed $N$ qubits interacting with a single
quantized mode of the bosonic field. Our aim is now to discuss the
interaction of a system of $N$ identical qubits with energy gap
$(\Omega=\omega_{2}-\omega_{1})$, with an infinite number of
harmonic oscillators which defines the reservoir. Let
$a_{k}^{\dagger}$ and $a_{k}$ be the creation and annihilation
operators of the $k-th$ harmonic oscillator of frequency
$\omega_{k}$. These creation and annihilation operators satisfy
the standard commutation relations $[a_{k}^{\dagger},
a_{q}]=\delta_{kq}$, $[a_{k}, a_{q}]=0$ and also
$[a_{k}^{\dagger}, a_{q}^{\dagger}]=0$. The total Hamiltonian,
i.e., the Hamiltonian of the combined system of the reservoir and
the $N$ identical qubits interacting with the reservoir reads
\begin{equation}
H=I_R\,\otimes\,\frac{\Omega}{2}\sum_{j=1}^{N}\,\sigma_{(j\,)}^z
+\sum_{k}\omega_{k}\,a_k^{\dagger}\,a_k\,\otimes\,I_S+
\frac{g}{\sqrt{N}}\sum_{j\,=1}^{N}\sum_{k}\Bigl(a_{k}\,\otimes\,
\sigma_{(j\,)}^{+}+a^{\dagger}_{k}\,\otimes\,
\sigma_{(j\,)}^{-}\Bigr)\, . \label{29}
\end{equation}

In the Eq. (\ref{29}) the first term in the right side is the free
Hamiltonian of of $N$ identical qubits, the second term is the
free harmonic oscillators reservoir Hamiltonian and finally the
third term is the interaction Hamiltonian between the reservoir
and the $N$ identical qubits. Note that we shift the zero of
energy for each qubits, as we did before, and we are assuming the
rotating-wave approximation, where $g(\sqrt{N})^{\,-1}$ is the
j-th qubit, k-th harmonic oscillator coupling constant. We also
can use a different interaction Hamiltonian to study the influence
of decoherence in quantum computers as was introduced by Di
Vicenzo \cite{di}. This author proposed the following model
describing a system of one qubit coupled to a reservoir of
harmonic oscillators:
\begin{eqnarray}
&&
I_{R}\,\otimes\,H_{S}+H_{R}\,\otimes\,I_{S}+H_{I}=\nonumber\\
&& I_{R}\,\otimes\frac{\Omega}{2}\,\sigma^z+\sum_{k}\omega_
{k}\,a_{k}^{\dagger}\,a_{k}\,\otimes\,I_{S}+ g\sum_{k}
\left(a_{k}^{\dagger}+ a_{k}\right)\,\otimes\,\sigma^z, \label{30}
\end{eqnarray}
where $\Omega$ is the usual energy level spacing of the qubit,
$a_{k}^{\dagger}$ and $a_{k}$ are respectively the boson creation
and annihilation operators associated to the harmonic oscillators.
Note the particular coupling between the reservoir and the qubit.
It is well known that this model allow for an exact analytic
solution and also exhibits the destruction of quantum coherence
without decay of population. There are two straightforward
generalizations for this model given by Eq. (\ref{30}). The first
one is to introduce $N$ identical qubits and the total Hamiltonian
for the composed system reads
\begin{eqnarray}
&&
I_{R}\,\otimes\,H_{S}+H_{R}\,\otimes\,I_{S}+H_{I}=\nonumber\\
&& I_{R}\,\otimes\,\frac{\Omega}{2}\sum_{j=1}^{N}
\,\sigma^{z}_{(j)}+\sum_{k}\omega_{k}\,a_{k}^{\dagger}\,a_{k}\,\otimes\,I_{S}+
\frac{g}{\sqrt{N}}\sum_{j\,=1}^{N}
\sum_{k}\,\left(a_{k}^{\dagger}+
a_{k}\right)\,\otimes\,\sigma^{z}_{(j\,)}  . \label{31}
\end{eqnarray}
Other generalization for the Di Vicenzo model is to introduce a
arbitrary mode-dependent coupling constant \cite{benatti}.
Therefore we have the following model describing a system of one
qubit coupled to a reservoir of harmonic oscillators:
\begin{eqnarray}
&&
I_{R}\,\otimes\,H_{S}+H_{R}\,\otimes\,I_{S}+H_{I}=\nonumber\\
&&
I_{R}\,\otimes\frac{\Omega}{2}\,\sigma^z+\sum_{k}\omega_{k}\,
a_{k}^{\dagger}\,a_{k}\,\otimes\,I_{S}+
\sum_{k} \left(\lambda_{k}\,a_{k}^{\dagger}+
\lambda_{k}^{*}\,a_{k}\right)\,\otimes\,\sigma^z, \label{gen}
\end{eqnarray}
Working in this line, the Hamiltonian describing a system of $N$
qubit coupled to a reservoir of harmonic oscillators reads
\cite{parma}
\begin{eqnarray}
&&
I_{R}\,\otimes\,H_{S}+H_{R}\,\otimes\,I_{S}+H_{I}=\nonumber\\
&& I_{R}\,\otimes\,\frac{\Omega}{2}\,
\sum_{j\,=1}^{N}\sigma^z_{(j\,)}+\sum_{k}\omega_{k}\,
a_{k}^{\dagger}\,a_{k}\,\otimes\,I_{S}+\frac{1}{\sqrt{N}}\sum_{j\,=1}^{N}
\sum_{k} \left(g_{j\,k}\,a_{k}^{\dagger}+
g_{j\,k}^{*}\,a_{k}\right)\,\otimes\,\sigma^z_{(j\,)},
\label{gen2}
\end{eqnarray}
where $g_{jk}$ describes the coupling between the j-th qubit with
the reservoir.

After this briefly discussion of many models describing the
two-level atoms-Bose field coupled system we would like to go back
to Eq.(\ref{29}). Going back to Eq. (\ref{29}), another
possibility is not assume the rotating-wave approximation in the
interaction Hamiltonian. Without the rotating-wave approximation,
the interaction Hamiltonian between the $N$ qubits and the
reservoir of harmonic oscillators reads
\begin{equation}
H_{I}=\frac{g}{\sqrt{N}}\sum_{j\,=1}^{N}\sum_{k} \left(
a_{k}+a_{k}^{\dagger}\right)\,\otimes\, \left(\sigma_{(j)}^{+}+
\sigma_{(j\,)}^{-}\right). \label{32}
\end{equation}
Our aim is to show that the generalized fermion Dicke model where
we are not assuming the rotating-wave-approximation also present a
phase transition and present the spectrum of the collective
bosonic excitations of the model. In the next section we will
study the Hilbert space for the fermion Dicke model. The main
points we wish to develop in the next section is the functional
integral for the generalized fermion Dicke model and also discuss
the non-analytic behavior of thermodynamic quantities at some
critical temperature.

\section{The Hilbert space for the fermion Dicke model}

\quad\, It should be emphasized that there are different routs to
proceed in the study of cooperative spontaneous emission of a
radiation field in the Dicke model. The first one is to introduce
collective atomic operators
\begin{equation}
J^{z}= \sum_{i=1}^N \sigma_{(i)}^{z}, \label{dif1}
\end{equation}
\begin{equation}
J^{+}= \sum_{i=1}^N \sigma_{(i)}^{+},
 \label{dif2}
\end{equation}
and
\begin{equation}
J^{-}= \sum_{i=1}^N \sigma_{(i)}^{-}.
 \label{dif3}
\end{equation}
These operators $J^{z}$, $J^{+}$ and $J^{-}$ obey the standard
angular momentum commutation relations. The Hilbert space of this
algebra is spanned by the Dicke states $|j\,,m\rangle$ which are
eigenstates of $J^{2}$ and $J^{z}$. This method allow us to
describe the $N$ two-level atoms by a single $(N+1)$ level system.
A second one method as we stressed before, is to change the atomic
pseudo-spin operators of the Dicke model by a linear combination
of Fermi operators to define the fermion Dicke model. The
dimensionality of the space where the combinations of Fermi
operators act are greater than the dimensionality of the space
where the pseudo-spin operators act, and therefore we have a
problem of the elimination of the superfluous states. We start
with the Hamiltonian of the Dicke model, $H_D$ given by
\begin{equation}
H_D=I_S\,\otimes\,\frac{\Omega}{2}\sum_{i=1}^N \sigma_{(i)}^{z}+
\omega_0\; b^{\dagger}\,b\,\otimes\,I_B+
\frac{g}{\sqrt{N}}\sum_{i=1}^N \Bigl(b\,\otimes\,\sigma_{(i)}^{+}+
b^{\dagger}\,\otimes\,\sigma_{(i)}^-\Bigr)\, . \label{33}
\end{equation}

Note that the $H_D$ Hamiltonian contains the pseudo-spin operators
used to obtain a second quantized version for the free two-level
atoms and also the creation and annihilation operators of the
single quantized mode of the bosonic field. As we discussed
before, the pseudo-spin operators $\sigma^z_{(i)}$,
$\sigma^+_{(i)}$ and $\sigma^-_{(i)}$, for the same or different
qubits obey the standard commutation relations
\begin{equation}
\left[\,\sigma_{(i)}^{+},\sigma_{(j)}^{-}\,\right]=\,\sigma_{(j)}^z\,\delta_{\,(i)}^{(j)}
, \label{n22}
\end{equation}
\begin{equation}
\left[\,\sigma_{(i)}^z,\sigma_{(j)}^{+}\,\right]=2\,\sigma_{(j)}^{+}\,\delta_{\,(i)}^{(j)}
, \label{n23}
\end{equation}
and finally
\begin{equation}
\left[\,\sigma_{(i)}^z,\sigma_{(j)}^{-}\,\right]=-2\,\sigma_{(j)}^{-}\,\delta_{\,(i)}^{(j)},
\label{n24}
\end{equation}
where $\delta_{\,(i)}^{(j)}$ is the Kronecker delta.

Let us define the Fermi raising and lowering operators
$\alpha^{\dagger}_{i}$, $\alpha_{i}$, $\beta^{\dagger}_{i}$ and
$\beta_{i}$, that satisfy the anti-commutator relations
$\alpha_{i}\alpha^{\dagger}_{j}+\alpha^{\dagger}_{j}\alpha_{i}
=\delta_{ij}$ and
$\beta_{i}\beta^{\dagger}_{j}+\beta^{\dagger}_{j}\beta_{i}
=\delta_{ij}$. We can also define the following bilinear
combination of Fermi operators, $\alpha^{\dagger}_{i}\alpha_{i}
-\beta^{\dagger}_{i}\beta_{i}$, $\alpha^{\dagger}_{i}\beta_{i}$
and finally $\beta^{\dagger}_{i}\alpha_{i}$. Note that
$\sigma^z_{(\,i)}$, $\sigma^+_{(\,i)}$ and $\sigma^-_{(\,i)}$ obey
the same commutation relations as the above presented bilinear
combination of Fermi operators. This suggests that we can change
the pseudo-spin operators of the Dicke model by the bilinear
combination of Fermi operators

\begin{equation}
\sigma_{(i)}^{z}\longrightarrow (\alpha_{i}^{\dagger}\alpha_{i}
-\beta_{i}^{\dagger}\beta_{i})\, , \label{34}
\end{equation}
\begin{equation}
\sigma_{(i)}^{+}\longrightarrow \alpha_{i}^{\dagger}\beta_{i}\, ,
\label{35}
\end{equation}
and finally
\begin{equation}
\sigma_{(i)}^{-}\longrightarrow \beta_{i}^{\dagger}\alpha_{i}\, .
\label{36}
\end{equation}
>From now on we use the usual notation instead of the notation
stressing the tensor product space of the total Hilbert space of
the system. With the  substitutions that we defined in the Eq.
(\ref{34}), Eq.(\ref{35}) and Eq.(\ref{36}) we shall call the
resulting Hamiltonian as the fermion Dicke model, i.e., $H_F$. The
Hamiltonian of the fermion Dicke model can be written as
\begin{equation}
H_F=\omega_0\; b^{\dagger}b+ \frac{\Omega}{2}\sum_{i=1}^N
\Bigl(\alpha_i^{\dagger}\alpha_i -\beta_i^{\dagger}\beta_i\Bigr) +
\frac{g}{\sqrt{N}}\sum_{i=1}^N \Bigl(b\,\alpha_i^{\dagger}\beta_i
\,+\, b^{\dagger}\, \beta_i^{\dagger}\alpha_i \Bigr)\, .
\label{37}
\end{equation}
Note that in the Eq. (\ref{37}) we are still adopting the
rotating-wave approximation. We can also define the fermion Dicke
model, without assume the rotating-wave approximation. The
interaction Hamiltonian $H_{I}$ of the single quantized mode of
the bosonic field in the presence of the $N$ qubits without the
rotating-wave approximation is given by
\begin{equation}
H_{I}=\frac{g}{\sqrt{N}} \sum_{i=1}^N \Bigl(b^{\dagger}+\,
b\Bigr)\Bigl(\alpha_i^{\dagger}\beta_i\,+\,
\beta_i^{\dagger}\alpha_i\Bigr)\, . \label{38}
\end{equation}
In effect, this interaction Hamiltonian includes four kind of
processes that correspond to the absorption or emission of a
mode-excitation of the field with the transition of the qubit from
its lower (upper) to its upper (lower) state. Therefore the
interaction Hamiltonian includes virtual processes or the vacuum
fluctuations contributions. A simple generalization of Eq.
(\ref{38}) that we are adopting in the paper is to introduce two
different coupling constants, the first one for the rotating-wave
terms and another one for the counter-rotating terms. We are  The
advantage of this method is that it is possible to identify the
contributions of the virtual processes and real ones in the
formation of the bosonic condensate.

After this discussion, let us analyze the Hilbert space of the
fermion Dicke model. The fermion Hilbert space for each atom is
four dimensional. The Hilbert space for the i-th atom is generated
by the following term vectors. The first one is the vacuum state
\begin{equation}
\phi_0=|\,0,0\rangle_{i}\, .
\label{39}
\end{equation}
Next, we can define more two vectors applying the operators
$\alpha^{\dagger}_i$ and $\beta^{\dagger}_i$ on the vacuum state,
i.e., $\alpha^{\dagger}_i \phi_0$ and $\beta^{\dagger}_i \phi_0$.
Therefore we have
\begin{equation}
\alpha^{\dagger}_{i}|\,0,0\rangle_{i}=|\,1,0\rangle_{i}
\label{40}
\end{equation}
and
\begin{equation}
\beta^{\dagger}_{i}|\,0,0\rangle_{i}=|\,0,1\rangle_{i}\, .
\label{41}
\end{equation}
Finally, the bilinear combination of the Fermi operators
$(\alpha_{i}^{\dagger}\beta^{\dagger}_{i})$ acting on the vacuum
state yields
\begin{equation}
\alpha^{\dagger}_{i} \beta^{\dagger}_{i}
|\,0,0\rangle_{i}=|\,1,1\rangle_{i}\, .
\label{42}
\end{equation}
The vectors $|\,1,0\rangle_i$ and $|\,0,1\rangle_i$ generate a
two-dimensional subspace, which is characterized by the following
condition with the bilinear combination of Fermi operators
\begin{equation}
\Bigl(\alpha^{\dagger}_{i} \alpha_{i} + \beta^{\dagger}_{i}
\beta_{i}\Bigr) |\,0,1\rangle_{i}=|\,0,1\rangle_{i}
\label{43}
\end{equation}
and
\begin{equation}
\Bigl(\alpha^{\dagger}_{i} \alpha_{i} + \beta^{\dagger}_{i}
\beta_{i}\Bigr) |\,1,0 \rangle_{i}=|\,1,0 \rangle_{i}\, .
\label{44}
\end{equation}
\quad We define $N_{i}=(\alpha^{\dagger}_{i} \alpha_{i} +
\beta^{\dagger}_{i} \beta_{i})$ as the fermion number operator
acting in the Hilbert space corresponding to the i-th atom. With
the four dimensional Hilbert space we can construct a physical
Hilbert space and a nonphysical one generated by the vectors
$|\,\Psi\rangle_i$ and $|\,\Phi\rangle_i$ respectively. Therefore
we have
\begin{equation}
c_{\,1}^{\,i} |\,0,1\rangle_{i} + c_{\,2}^{\,i}
|\,1,0\rangle_{i}=|\,\Psi\rangle_{i}
\label{45}
\end{equation}
and
\begin{equation}
d_{\,1}^{\,i} |\,0,0\rangle_{i} + d_{\,2}^{\,i}
|\,1,1\rangle_{i}=|\,\Phi\rangle_{i}\,.
\label{45a}
\end{equation}

Next we will obtain a formula which connect the partition function
of spin Dicke model to the partition function of the fermion Dicke
model. Using the definition of the fermion number operator of all
qubits, and that $N_i=\bigl(\alpha^{\dagger}_i \alpha_i +
\beta^{\dagger}_i \beta_i\bigr)$ we have
\begin{equation}
N=N_{i} + \sum_{j\neq i} N_{j}\, .
\label{46}
\end{equation}
Since the number operator and the Hamiltonian operator commute and
using the same notation we have
\begin{equation}
H_{F}=H_{i} + \sum_{j\neq i} H_{j}\, .
\label{47}
\end{equation}
Using the fact that
\begin{equation}
H_{i}|\,0,0\rangle _{i}=H_{i}|\,1,1\rangle _{i}= 0\, ,
\label{48}
\end{equation}
we have that
\begin{equation}
H_{i}|\,\Phi \rangle _{i} =0\, ,
\label{49}
\end{equation}
where $|\,\Phi\rangle _i$ is the general vector state in the
nonphysical subspace of the $i$-th qubit. The trace over the
nonphysical states of the $i$-th qubit vanishes. Therefore we have
\begin{equation}
_{i}\langle \Phi\,| \exp\Biggl[-\beta\biggl(H_{i}+\sum_{j\neq i}
H_{j}+\frac{i\pi}{2\beta}N\biggr)\Biggr] |\,\Phi\rangle_{i}=
\label{50}
\end{equation}
\begin{equation}
_{i}\langle \Phi\,| \exp\Biggl[-\beta\biggl(H_{i}+\sum_{j\neq i}
H_{j}\biggr)+\frac{i\pi}{2}\biggl(N_{i}+\sum_{j\neq
i}N_{j}\biggr)\Biggr] |\,\Phi\rangle_{i}=0\, . \label{51}
\end{equation}
Therefore we have
\begin{equation}
Tr \exp\Biggl[{-\beta
\biggl(H_{F}+\frac{i\pi}{2\beta}N\biggr)}\Biggr]=(-i)^N Tr_{phys}
\exp[{-\beta H_{F}}]=(-i)^N Tr \exp[{-\beta H_{\sigma}}]\, ,
\label{52}
\end{equation}
and we can present the relation between the partition function of
the spin Dicke model to the partition function of the fermion
Dicke model. A simple formula is given by
\begin{equation}
Tr\, \exp\,[{-\beta H_{\sigma}}]=i^N Tr\,
\exp\,\Biggl[{\biggl(-\beta H_F +\frac{i\pi}{2}N\biggr)}\Biggr]\,
. \label{53}
\end{equation}
According the Eq. (\ref{53}), to study the spin Dicke model we can
use the fermion Hamiltonian $H_F$ adding $\frac{i\pi N}{2\beta}$,
i.e. the fermion number operator with pure imaginary coefficient
(chemical potential). All the standard diagrammatic technique for
the Fermi system can be generated using the Fourier representation
for the Green function given by
\begin{equation}
G^{\,0}(\omega_{F})=\frac{1}{i\omega_F -\epsilon +\mu} =
\frac{1}{i\omega_F -\epsilon - \frac{i\pi}{2\beta}}\, ,
\label{53a}
\end{equation}
where $\omega_F= \frac {2\pi}{\beta}(n+1/2)$ is the fermion
Matsubara frequency.
After the above discussion we can consider the problem of define
the partition function of the fermion Dicke model defined by
$Z_F$.

\section{The functional integral for the generalized fermion Dicke model}

\quad After the above discussion we can consider the problem of
define the partition function $Z_F$ of the generalized fermion
Dicke model. First let us define the Euclidean action $S$ of this
model, which describes a single quantized mode of the field and
the ensemble of $N$ identical qubits. The Euclidean action $S$ is
given by
\begin{equation}
S=\int_0^{\beta} d\tau \biggl(b^*(\tau)\frac{\partial}{\partial
\tau} b(\tau)+ \sum_{i=1}^{N}
\Bigl(\alpha^*_i(\tau)\frac{\partial}{\partial \tau}\alpha_i(\tau)
+\beta^*_i (\tau)\frac{\partial}{\partial
\tau}\beta_i(\tau)\Bigr)\biggr) -\int_0^{\beta}d\tau H_{F}(\tau)\,
\label{66}
\end{equation}
where $H_{F}$ is the full Hamiltonian for the generalized fermion
Dicke model given by
\begin{eqnarray}
H_{F}\,=\,\omega_{0}\,b^{\,*}(\tau)\,b(\tau)\,+
\,\frac{\Omega}{2}\,\displaystyle\sum_{i\,=\,1}^{N}\,
\biggl(\alpha^{\,*}_{\,i}(\tau)\,\alpha_{\,i}(\tau)\,-
\,\beta^{\,*}_{\,i}(\tau)\beta_{\,i}(\tau)\biggr)\,+
\nonumber\\
+\,\frac{g_{\,1}}{\sqrt{N}}\,\displaystyle\sum_{i\,=\,1}^{N}\,
\biggl(\alpha^{\,*}_{\,i}(\tau)\,\beta_{\,i}(\tau)\,b(\tau)\,+
\alpha_{\,i}(\tau)\,\beta^{\,*}_{\,i}(\tau)\,b^{\,*}(\tau)\,\biggr)\,+
\nonumber\\
+\,\frac{g_{\,2}}{\sqrt{N}}\,\displaystyle\sum_{i\,=\,1}^{N}\,
\biggl(\alpha_{\,i}(\tau)\,\beta^{\,*}_{\,i}(\tau)\,b(\tau)\,+
\,\alpha^{\,*}_{\,i}(\tau)\,\beta_{\,i}(\tau)\,b^{\,*}(\tau)\biggr).
\label{66a}
\end{eqnarray}
Note that we are introducing two coupling constants, $g_{1}$ and
$g_{2}$, for the rotating and anti-rotating wave terms
respectively. As we discussed before, the main reason for this is
that we are interested to identify the contribution of the real
and virtual processes in the phase transition with the formation
of the condensate.  Let us define the formal quotient of two
functional integrals, i.e., the partition function of the
generalized fermion Dicke model and the partition function of the
free fermion Dicke model. Therefore we are interested in calculate
the following quantity
\begin{equation}
\frac{Z_{F}}{Z_{F_{0}}}=\frac{\int [d\eta]\,e^{\,S}}{\int
[d\eta]\,e^{\,S_{0}}}\, , \label{65}
\end{equation}
where $S=S(b,b^*,\alpha,\alpha^{\dagger},\beta,\beta^{\dagger})$
is the Euclidean action of the generalized fermion Dicke model
given by Eq. (\ref{66}),
$S_0=S_{0}(b,b^*,\alpha,\alpha^{\dagger},\beta,\beta^{\dagger})$
is the free Euclidean action for the free single bosonic mode and
the free qubits, i.e., the expression of the complete action $S$
taking $g_1=g_2=0$ and finally $[d\eta]$ is the functional
measure.
The functional integrals involved in Eq. (\ref{65}), are
functional integrals with respect to the complex functions
$b^*(\tau)$ and $b(\tau)$ and Grassmann Fermi fields
$\alpha_i^*(\tau)$, $\alpha_i(\tau)$, $\beta_i^*(\tau)$ and
$\beta_i(\tau)$. Since we are using thermal equilibrium boundary
conditions, in the imaginary time formalism, the integration
variables in Eq. (\ref{65}) obey periodic boundary conditions for
the Bose field, i.e., $b(\beta)=b(0)$ and anti-periodic boundary
conditions for Fermi fields i.e., $\alpha_i(\beta)=-\alpha_i(0)$
and $ \beta_i(\beta)=-\beta_i(0)$.

The free action for the single mode bosonic field $S_{0}(b)$ is
given by
\begin{equation}
S_{0}(b) = \int_{0}^{\beta} d\tau \biggl(b^{*}(\tau)
\frac{\partial b(\tau)}{\partial \tau} -
\omega_{0}\,b^{*}(\tau)b(\tau)\biggr)\, . \label{67}
\end{equation}
Then we can write the action $S$ of the generalized fermion Dicke
model, given by Eq. (\ref{66}), using the free action for the
single mode bosonic field $S_{0}(b)$ given by Eq. (\ref{67}), plus
an additional term that can be expressed in a matrix form.
Therefore the total action $S$ can be written as
\begin{equation}
S = S_{0}(b) +  \int_{0}^{\beta} d\tau\,\sum_{i=1}^{N}\,
\rho^{\dagger}_{i}(\tau)\, M(b^{*},b)\,\rho_{i}(\tau)\, ,
\label{68}
\end{equation}
where $\rho_{\,i}(\tau)$ is a column matrix given in terms of
fermion field operators
\begin{eqnarray}
\rho_{\,i}(\tau) &=& \left(
\begin{array}{c}
\beta_{\,i}(\tau) \\
\alpha_{\,i}(\tau)
\end{array}
\right),
\nonumber\\
\rho^{\dagger}_{\,i}(\tau) &=& \left(
\begin{array}{cc}
\beta^{*}_{\,i}(\tau) & \alpha^{*}_{\,i}(\tau)
\end{array}
\right) \label{69a}
\end{eqnarray}
and the matrix $M(b^{*},b)$ is given by
\begin{equation}
M(b^{*},b) = \left( \begin{array}{cc}
\partial_{\tau} + \Omega/2 & (N)^{-1/2}\,\biggl(g_{1}\,b^{*}\,(\tau) + g_{2}\,b\,(\tau)\biggr)\\
(N)^{-1/2}\,\biggl(g_{1}\,b\,(\tau) + g_{2}\,b^{*}\,(\tau)\biggr)
&
\partial_{\tau} - \Omega/2
\end{array} \right)\,.
\label{69b}
\end{equation}
These fields $b(\tau)$, $\alpha_{i}(\tau)$ and $\beta_{i}(\tau)$
can be written as a Fourier expansion. Therefore we have
\begin{equation}
b(\tau) = \beta^{-1/2} \sum_{\omega} b(\omega) e^{i\omega \tau}\,
, \label{69c}
\end{equation}
and
\begin{equation}
\rho_{i}(\tau) = \beta^{-1/2} \sum_{p} \rho_{i}(p) e^{ip \tau}\, .
\label{71}
\end{equation}
Since the field $b(\tau)$ obeys periodic boundary conditions, and
the fields $\alpha_{i}(\tau)$ and $\beta_{i}(\tau)$ obey
anti-periodic boundary conditions, we have that $\omega =
\frac{2\pi n}{\beta}$ and $p=\frac{(2n+1)\pi}{\beta}$, where they
are the boson and fermion Matsubara frequencies respectively.
Substituting the Fourier expansions in the free action given by
Eq. (\ref{67}) we get
\begin{equation}
S_{0}(b) = \sum_{\omega} (i\omega -
\omega_{0})b^{*}(\omega)b(\omega). \label{72}
\end{equation}
Using the same procedure in the Eq. (\ref{68}) allow us to write
\begin{equation}
S = S_{0}(b) + \sum_{p,\,q} \sum_{i=1}^{N}\,
\rho^{\dagger}_{i}(p)\, M_{p\,q}(b^{*},b)\,\rho_{i}(q)\, ,
\label{72b}
\end{equation}
where the matrix $M_{p\,q}(b^{*},b)$ is given by
\begin{equation}
M_{p\,q}(b^{*},b) = \left( \begin{array}{cc}
(ip + \Omega/2)\delta_{p\,q} & (N\beta)^{-1/2}\biggl(g_{1}\,b^{*}(q-p) + g_{2}\,b(p-q)\biggr)\\
(N\beta)^{-1/2}\biggl(g_{1}\,b(p-q) + g_{2}\,b^{*}(q-p)\biggr) &
(ip - \Omega/2)\delta_{p\,q}
\end{array} \right)\, .
\label{73}
\end{equation}
Using the above results, the ratio between the two functional
integrals $Z$ and $Z_{0}$, i.e., $\frac{Z}{Z_{0}}$ is given by
\begin{equation}
\frac{\int [d\eta(b)]\exp\biggl(\displaystyle\sum_{\omega}
(i\omega - \omega_{0})b^{*}(\omega)b(\omega)\biggr)\int
[d\eta(\rho)]\exp\biggl(\sum_{p,q} \sum_{i=1}^{N}\,
\rho^{\dagger}_{i}(p)\,
M_{p\,q}(b^{*},b)\,\rho_{i}(q)\biggr)}{\int [d\eta(b)]
\exp\biggl(\displaystyle\sum_{\omega}(i\omega -
\omega_{0})b^{*}(\omega)b(\omega)\biggr) \int
[d\eta(\rho)]\exp\biggl(\sum_{p,q} \sum_{i=1}^{N}\,
\rho^{\dagger}_{i}(p)\, M_{p\,q}(0,0)\,\rho_{i}(q)\biggr)}
\label{74}
\end{equation}
where the functional measures $[d\eta(b)]$ and $[d\eta(\rho)]$ in
the above equation are defined respectively by
\begin{equation}
[d\eta(b)]=\prod_{\omega}db(\omega)db^{*}(\omega) \label{75}
\end{equation}
and
\begin{equation}
[d\eta(\rho)]=\prod_{i,\,p}d\rho_{i}(p)d\rho^{\dagger}_{i}(p)\, .
\label{76}
\end{equation}
We need impose cutoffs over the boson and fermion Matsubara
frequencies on these measures. This procedure is necessary to be
sure that the ratio between the two functional integrals given by
$\frac{Z}{Z_{0}}$ does not diverge. After all , at the end, we
must take these cutoffs to infinity. In order to define the
effective action associated to the bosonic mode, we integrate out
the fermionic degrees of freedom. The integrals with respect to
the Fermi fields are Gaussian and we may integrate over these
Grassmann variables. This procedure yields
\begin{equation}
\int[d\eta(\rho)]\exp\biggl(\sum_{p,q}
\sum_{i=1}^{N}\,\rho^{\dagger}_{i}(p)\,
M_{p\,q}(b^{*},b)\,\rho_{i}(q)\biggr)= \,\det\,^{N}M(b^{*},b)\, ,
\label{77}
\end{equation}
where the matrix $M$ is a block matrix of the following form
\begin{equation}
M(b^{*},b) = \left( \begin{array}{cc}
(ip + \Omega/2)\,I & (N\beta)^{-1/2}\,Q^{\dagger}\\
(N\beta)^{-1/2}\,Q & (ip - \Omega/2)\,I
\end{array} \right)\,,
\label{77b}
\end{equation}
where $I$ is the identity matrix and the components of matrix $Q$
are
\begin{equation}
Q_{p\,q}=g_{1}\,b(p-q) + g_{2}\,b^{*}(q-p)\,.
\end{equation}
The following change of coordinates can simplify our calculations.
Let us change variables in the following way:
\begin{equation}
b(\omega)\rightarrow~\Biggl(\frac{\pi}{(\omega_{0} -
i\omega)}\Biggr)^{1/2}b(\omega) \label{78}
\end{equation}
and
\begin{equation}
b^{*}(\omega)\rightarrow\Biggr(\frac{\pi}{(\omega_{0} -
i\omega)}\Biggr)^{1/2}b^{*}(\omega)\, . \label{79}
\end{equation}
We must note that Eq. (\ref{79}) is not conjugate of Eq.
(\ref{78}). Nevertheless, is not difficult to justify this
transformation, if we introduce polar coordinates instead of
$b(\omega),\,b^*(\omega):b(\omega)=(\rho(\omega))^{1/2}e^{i\phi(\omega)},
b^*(\omega)=(\rho(\omega))^{1/2}e^{-i\phi(\omega)}$ and then
perform a complex rotation of the integration counter when
integrating with respect to $\rho(\omega):\rho(\omega)\rightarrow
\rho(\omega)[\pi/(\omega_0-i\omega)]^{1/2}$. Its easy to see that
after these changes of variables the denominator of the Eq.
(\ref{74}), turns out to be equal to unity
\begin{equation}
\int [d\eta(b)] \exp\biggl(-\pi
\sum_{\omega}b^{*}(\omega)b(\omega)\biggr) = 1\, , \label{80}
\end{equation}
so we can express the ratio $\frac{Z}{Z_{0}}$ by the integral
\begin{equation}
\frac{Z}{Z_{0}}=\int[d\eta(b)]\,\exp{\biggl(\,S_{\,eff}\,(b)\,\biggr)}\,,
\label{81}
\end{equation}
where $S_{\,eff}(b)$ is the effective action of the bosonic mode
which is given by
\begin{equation}
S_{\,eff}\,=\,-\,\pi\,\sum_{\omega}\,b^{\,*}(\omega)\,
b(\omega)\,+\,N\ln{det\,(I\,+\,A)}\,. \label{81a}
\end{equation}
The determinant in the above equation is given by
\begin{equation}
\det(I+A) = \det\biggl(M^{-1/2}(0,0)M(b^{*},b)M^{-1/2}(0,0)\biggr)
\label{82}
\end{equation}
and the matrix A is defined as follows
\begin{equation}
A\,=\,\left( \begin{array}{cc} 0 & B\\ -C & 0
\end{array} \right)\, .
\label{83}
\end{equation}
In the equation above the quantities $B$ and $C$ are matrices with
components given by
\begin{equation}
B_{p\, q}=\Biggl(\frac{\pi}{\beta
N}\Biggr)^{\frac{1}{2}}\Biggl(ip+\frac{\Omega}{2}
\Biggr)^{-\frac{1}{2}}\Biggl(\,\frac{g_{\,1}\,b^{*}\,(q-p)}{\sqrt{\omega_{0}-i(q-p)}}
+\frac{g_{\,2}\,b\,(p-q)}{\sqrt{\omega_{0}-i(p-q)}}\Biggr)\,\Biggl(iq-\frac{\Omega}{2}
\Biggr)^{-\frac{1}{2}} \label{83a}
\end{equation}
and
\begin{equation}
C_{p\, q}=-\,\Biggl(\frac{\pi}{\beta
N}\Biggr)^{\frac{1}{2}}\Biggl(ip-\frac{\Omega}{2}
\Biggr)^{-\frac{1}{2}}\Biggl(\,\frac{g_{\,1}\,b\,(p-q)}{\sqrt{\omega_{0}-i(p-q)}}
+\frac{g_{\,2}\,b^{*}\,(q-p)}{\sqrt{\omega_{0}-i(q-p)}}\Biggr)\,\Biggl(iq+\frac{\Omega}{2}
\Biggr)^{-\frac{1}{2}} . \label{83b}
\end{equation}
In Eq. (\ref{81}) we may go to the limit $\omega_B, \,\omega_F
\rightarrow \infty$ and the instead of a formal quotient of two
infinite functional integrals we shall have only one finite
functional integral. This representation turns out to be very
useful for obtaining the asymptotic formula for $Z/Z_0$ at large
$N$. There exists only one stationary phase point at
$\beta^{-1}>\beta^{-1}_c$. If $\beta^{-1}<\beta^{-1}_c$, we have a
circle of a stationary phase
$|\,b(0)|\,^2=\rho_0,\,b(\omega)=b^*(\omega)=0$, if $\omega\neq
0$. There also exists an interpolation formula between these
asymptotes. The presence of degenerate vacua is a feature of
states with spontaneous symmetry breaking. As we will see, gapless
excitation will appear.

We shall investigate the integral given by Eq. (\ref{81}) for
temperatures that satisfy  $\beta^{-1}>\beta^{-1}_c$. First of all
let us show that this integral converges. We use the estimate
\begin{equation}
|\det(I+A)|\leq \exp\biggl(\mbox{Re}\,(tr A)+\frac{1}{2}\,tr (A
A^{\dagger})\biggr)\,.
\end{equation}
where $\mbox{Re}\,(tr A)$ means the real part of $tr A$. The
matrix $A$ have the form given by Eq. (\ref{83}). Therefore we
find that $tr A=0$ and $tr (A A^{\dagger})=tr (B B^{\dagger})+tr
(C C^{\dagger})$. Therefore, we obtain the estimate
\begin{eqnarray}
\frac{Z}{Z_{0}}&\leq&\int[d\eta(b)]\,\exp{\biggl(\,-\,\pi\,
\sum_{\omega}\,b^{\,*}(\omega)\,
b(\omega)\,+\,N\,tr (B B^{\dagger})\,+\,N\,tr (C C^{\dagger})\,\biggr)}\,,\nonumber\\
&\leq&\int[d\eta(b)]\,\exp\Biggl(-\,\pi\,\sum_{\omega}\,b^{\,*}
(\omega)\,\biggl(1\,-\,a_0(\omega)\,\biggr)\,b(\omega)
\,+\nonumber\\
&
&+\,\pi\,\sum_{\omega}\,\biggl(b(\omega)\,c_0(\omega)\,b(\,-\,\omega)\,+\,
b^{\,*}(\omega)\,c_0(\omega)\,b^{\,*}(-\,\omega)\,\biggr)\,\Biggr)\,,
\label{83d}
\end{eqnarray}
where the $a_{0}(\omega)$ and $c_{0}(\omega)$ are given
respectively by
\begin{equation}
a_{0}(\omega)\,=\,\frac{g_{\,1}^{\,2}\,+\,g_{\,2}^{\,2}}{\beta\,(\omega^{\,2}_{\,0}\,
+\,\omega^{\,2})^{\,1/2}}\,\displaystyle\sum_{p\,-\,q\,=\,\omega}\,\frac{1}{(\frac{\Omega^{\,2}}{4}\,+\,
q^{\,2})^{\,1/2}}\,\frac{1}{(\frac{\Omega^{\,2}}{4}\,+\,
p^{\,2})^{\,1/2}},
\end{equation}
and
\begin{equation}
c_{0}(\omega)\,=\,\frac{\omega_0\,g_{\,1}\,g_{\,2}}{\beta\,(\omega^{\,2}_{\,0}\,
+\,\omega^{\,2})}\,\displaystyle\sum_{p\,-\,q\,=\,\omega}\,\frac{1}{(\frac{\Omega^{\,2}}{4}\,+\,
q^{\,2})^{\,1/2}}\,\frac{1}{(\frac{\Omega^{\,2}}{4}\,+\,
p^{\,2})^{\,1/2}}.
\end{equation}
Using the measure given in Eq. (\ref{75}), we have that $
\frac{Z}{Z_{0}}\leq F$, where $F=F_1\,F_2$ and where $F_1$ and
$F_2$ are given by
\begin{eqnarray}
& &F_1=\int\,db(0)db^{*}(0)\nonumber\\
& &\exp\Biggl[-\,\pi\,b^{\,*}
(0)\,\Bigl(1\,-\,a_0(0)\,\Bigr)\,b(0)\,+\,\pi\,
\biggl(b(0)\,c_0(0)\,b(0)\,+\,
b^{\,*}(0)\,c_0(0)\,b^{\,*}(0)\,\biggr)\Biggr]
\end{eqnarray}
and
\begin{eqnarray}
& &F_2=\int\,\prod_{\omega>\,0}db(\omega)db^{*}(\omega)
db(-\omega)db^{*}(-\omega)\nonumber\\
& &\exp\Biggl[-\,\pi\,\sum_{\omega>\,0}\,b^{\,*}
(\omega)\,\Bigl(1\,-\,a_0(\omega)\,\Bigr)\,b(\omega)\,-\,\pi\,\sum_{\omega>\,0}\,b^{\,*}
(-\omega)\,\Bigl(1\,-\,a_0(\omega)\,\Bigr)\,b(-\omega)\,
+\nonumber\\
& &2\,\pi\,\sum_{\omega>\,0}\,\biggl(b(\omega)
\,c_0(\omega)\,b(\,-\,\omega)\,+\,
b^{\,*}(\omega)\,c_0(\omega)\,b^{\,*}(-\,\omega)\,\biggr)\,\Biggr]\,.
\end{eqnarray}
Note that in the case of the generalized fermion Dicke model we
obtained a Gaussian integral that mixtures positive with negative
frequencies. A straightforward calculation gives that the ratio
$\frac{Z}{Z_{0}}$ obeys the following inequality
\begin{eqnarray}
\frac{Z}{Z_{0}}&\leq&\Biggl[\,\biggl(1\,-\,a_0(0)\,+
\,2\,c_0(0)\biggr)\,\biggl(1\,-\,a_0(0)\,-\,2\,c_0(0)\,\biggr)\,\Biggr]^{\,-\,1/2}\,
\nonumber\\
&&\prod_{\,\omega\,>\,0}\,\Biggl[\,\biggl(1\,-\,a_0(\omega)\,+
\,2\,c_0(\omega)\biggr)\,\biggl(1\,-\,a_0(\omega)\,-\,2\,c_0(\omega)\,\biggr)\,
\Biggr]^{\,-\,1}\,. \label{83c}
\end{eqnarray}
In a similar way like Popov and Fedotov \cite{popov} proved, for
the case of rotating wave approximation, we have that,
$0<a_0(\omega)\,+ \,2\,c_0(\omega)<a_0(0)\,+\,2\,c_0(0)$ and
$a_0(0)\,+\,2\,c_0(0)=O(\omega^{-2}\ln\,\omega)$. Therefore if
$a_0(0)\,+\,2\,c_0(0)<1$, then Eq. (\ref{83c}) guarantees
convergence of the expression $\frac{Z}{Z_0}$. The condition
$a_0(0)\,+\,2\,c_0(0)=1$ is the equation for the transition
temperature, then we have
\begin{equation}
a_0(0)\,+\,2\,c_0(0)\,=\,\frac{(\,g_{\,1}+g_{\,2}\,)^{\,2}}
{\Omega\,\omega_{0}}\,\tanh\biggl(\frac{\beta_{c}\,\Omega}{4}\biggr)\,=1\,.
\end{equation}
The inverse of the critical temperature $\beta_{c}$ is given by
\begin{equation}
\beta_{c} =
\frac{4}{\Omega}\tanh^{-1}\biggl(\frac{\Omega\,\omega_{0}}{(\,g_{\,1}+
g_{\,2}\,)^{\,2}}\biggr)\,.
\end{equation}
Note that there is a quantum phase transition where the coupling
constants $g_{1}$ and $g_{2}$ satisfy
$g_{1}+g_{2}=(\omega_{0}\,\Omega)^{\frac{1}{2}}$. For larger
values for ($g_{1}+g_{2}$) the system enters in a superradiant
phase.

To calculate the asymptotic behavior of the functional integrals
at temperatures that satisfy $\beta^{-1}>\beta^{-1}_{c}$, we can
do the following approximation
\begin{equation}
\det\,^{N}(I+A) = \det\,^{N}(I+BC)\rightarrow
\exp\biggl(N\,tr(BC)\biggr)\, . \label{84}
\end{equation}
This substitute can be done and we can estimate the error if we
divide all the functional space into two domains $C_{1}$ and
$C_{2}$
\begin{equation}
tr\Bigl[(BC)(BC)^{\dagger}\Bigr]\leq(4N)^{-1} \mapsto C_{1}\, ,
\label{85}
\end{equation}
\begin{equation}
tr\Bigl[(BC)(BC)^{\dagger}\Bigr]\geq(4N)^{-1} \mapsto C_{2}\, .
\label{86}
\end{equation}
Denoting
\begin{equation}
K_{N} = \det\,^{N}(I+A) - \exp\biggl(Ntr(BC)\biggr)\, , \label{87}
\end{equation}
for the ratio $\frac{Z}{Z_{0}}$, we have the following identity
\begin{eqnarray}
\frac{Z}{Z_{0}} =
\int[d\eta(b)]\exp\biggl(-\pi\sum_{\omega}b^{*}(\omega)b(\omega)+N\,tr(BC)\biggr)+
\nonumber\\
+ \int_{C_{1}}[d\eta(b)]K_{N}\exp\biggl(-\pi\sum
b^{*}(\omega)b(\omega)\biggr)+
\nonumber\\
+\int_{C_{2}}[d\eta(b)]K_{N}\exp\biggl(-\pi\sum
b^{*}(\omega)b(\omega)\biggr)\, . \label{88}
\end{eqnarray}
The first integral of the above equation is Gaussian, let us
define it by $I_{0}$. We use the Eq. (\ref{83a}) and the Eq.
(\ref{83b}) in order to calculate the trace of BC, i.e., $tr(BC)$
which is present in the expression $I_{0}$. A simple calculation
gives
\begin{eqnarray}
I_{0}\,=\,\int\,[d\eta(b)]\,\exp\Biggl(-\,\pi\,\sum_{\omega}\,b^{\,*}
(\omega)\,\biggl(1\,-\,a(\omega)\,\biggr)\,b(\omega) \,+
\nonumber\\
+\,\pi\,\sum_{\omega}\,\biggl(b(\omega)\,c(\omega)\,b(\,-\,\omega)\,+\,
b^{\,*}(\omega)\,c(\omega)\,b^{\,*}(-\,\omega)\,\biggr)\,\Biggr)\,,
\label{89}
\end{eqnarray}
where $a(\omega)$ and $c(\omega)$ of above equation are given
respectively by
\begin{equation}
a(\omega)\,=\,\Biggl(\frac{g_{\,1}^{\,2}\,(\Omega
-i\omega)^{\,-1}+\,g_{\,2}^{\,2}\,(\Omega\,+\,
i\omega)^{\,-1}}{(\omega_{0}\,-\,i\,\omega)}
\Biggr)\,\tanh{\biggl(\,\frac{\beta\,\Omega}{4}\,\biggr)}\,
\label{90}
\end{equation}
and
\begin{equation}
c(\omega)\,=\,\Biggl(\frac{g_{\,1}\,g_{\,2}\,\Omega}{(\omega_{\,0}^{\,2}\,+
\,\omega^{\,2})^{\,1/2}\,(\Omega^{\,2}\,+\,\omega^{\,2})}\Biggr)
\,\tanh{\,\biggl(\frac{\beta\,\Omega}{4}\biggr)}.
 \label{90a}
\end{equation}
Note that to recover the result obtained by Popov and Fedotov
\cite{pri} \cite{popov} we have only to assume $g_{2}=0$. In this
case, we have that $c(\omega)=0$, which simplify the integration
over the mode of the bosonic field in Eq. (\ref{89}). Making the
integration we obtain that $I_{\,0}$ is given by
\begin{equation}
I_{0}=\prod_{\omega}\biggl(\,1-a(\omega)\,\biggr)^{\,-1}.
\end{equation}
After this observation lets go back to the general case, where
$g_{1}$ and $g_{2}$ take arbitrary values. The expression $I_0$
given in Eq. (\ref{89}) is a Gaussian integral, this expression is
similar to the integral given in Eq. (\ref{83d}), so following the
same steps we get that
\begin{equation}
I_{\,0}\,=\,I_{\,0}(\omega=0)\,\prod_{\,\omega\,>\,0}\,\Biggl[c(\omega)^
{\,2}\,-\,\biggl(1\,-\,a(\omega)\,\biggr)\,
\biggl(1\,-\,a(\,-\omega)\,\biggr)\,\Biggr]^{\,-\,1}\,, \label{95}
\end{equation}
where $I_{\,0}(\omega=0)$ is the contribution of the condensate
given by
\begin{equation}
I_{\,0}(\omega=0)\,=\,\Biggl[\,\biggl(1\,-\,a(0)\,+
\,2\,c(0)\biggr)\,\biggl(1\,-\,a(0)\,-\,2\,c(0)\,\biggr)\,\Biggr]^{\,-\,1/2}.
\label{95a}
\end{equation}
It is possible to estimate the error of $I_{0}$ which is given by
the two last terms of Eq. (\ref{88}). For details see the
reference \cite{pop2}. The errors depend on order $N^{-1}$.
Therefore $\frac{Z}{Z_{\,0}}$ can be written as
\begin{eqnarray}
\frac{Z}{Z_{\,0}}&=& \Biggl[\,\biggl(1\,-\,a(0)\,+
\,2\,c(0)\biggr)\,\biggl(1\,-\,a(0)\,-\,2\,c(0)\,\biggr)\,\Biggr]^{\,-\,1/2}\,
\\
\nonumber
&&\prod_{\,\omega\,>\,0}\,\Biggl[\,\biggl(1\,-\,a(\omega)\,\biggr)\,
\biggl(1\,-\,a(\,-\omega)\,\biggr)\,-\,c^{\,2}(\omega)\,\Biggr]^{\,-\,1}\,+
\\
\nonumber
\\
\nonumber &&+O(N^{\,-1})\,, \label{95}
\end{eqnarray}
therefore in the limit ($N\rightarrow \infty$) the equality
$\frac{Z}{Z_0}=I_0$ is a good approximation. To find the
collective excitation spectrum we have to use the equation
\begin{equation}
c^{\,2}(\omega)\,-\,\biggl(1\,-\,a(\omega)\,\biggr)\,
\biggl(1\,-\,a(\,-\omega)\,\biggr)\,=0\, , \label{105}
\end{equation}
and making the analytic continuation $(i\omega \rightarrow E)$, we
obtain the following equation
\begin{eqnarray}
&&1\,=\,-\Biggl[\frac{g_{\,1}^{\,4}\,+\,g_{\,2}^{\,4}}
{(\omega_{\,0}^{\,2}\,-\,E^{\,2})\,(\Omega^{\,2}\,-\,E^{\,2})}\Biggr]\,
\tanh^{\,2}\biggl(\frac{\beta\,\Omega}{4}\biggr)\,+
\nonumber\\
\nonumber\\
&&-\Biggl[\frac{g_{\,1}^{\,2}\,g_{\,2}^{\,2}}{(\omega_{\,0}^{\,2}
\,-\,E^{\,2})}\Biggl(\frac{1}{(\Omega\,-E)^{\,2}}\,+
\,\frac{1}{(\Omega\,+\,E)^{\,2}}\,-\,\frac{4\,\Omega^{\,2}}
{(\Omega^{\,2}\,-\,E^{\,2})^{\,2}}\Biggr)\,\Biggr]\,
\tanh^{\,2}\Biggl(\frac{\beta\,\Omega}{4}\Biggr)\,+
\nonumber\\
\nonumber\\
&&+\,\Biggl[\frac{g_{\,1}^{\,2}(\,\Omega\,-\,E\,)^{\,-1}\,+
\,g_{\,2}^{\,2}(\,\Omega\,+\,E\,)^{\,-1}}{(\omega_{0}\,-\,E)}\,+
\frac{g_{\,1}^{\,2}(\,\Omega\,+\,E\,)^{\,-1}\,+
\,g_{\,2}^{\,2}(\,\Omega\,-\,E\,)^{\,-1}}{(\omega_{0}\,+\,E)}\Biggr]
\,\tanh\Biggl(\frac{\beta\,\Omega}{4}\Biggr).\nonumber\\
\label{105}
\end{eqnarray}
Solving the above equation for the case $\beta^{-1}=\beta^{-1}_c$,
we find the following roots
\begin{equation}
E_{\,1}\,=\,0
\label{106}
\end{equation}
and
\begin{equation}
E_{\,2}\,=\,\Biggl(\,\frac{g_{\,1}\,(\Omega\,+\,\omega_{\,0})^{\,2}\,+\,
g_{\,2}\,(\Omega\,-\,\omega_{\,0})^{\,2}}{(g_{\,1}\,+\,g_{\,2})}\,\Biggr)^{\,1/2}\,.
\label{107}
\end{equation}
Its low energy state of excitation is a Goldstone mode. Now, let
us present the critical temperature and the spectrum of the
collective bosonic excitations of the model with the rotating-wave
approximation, where $g_{1}\neq 0$ and $g_{2}=0$.
The result obtained by Popov and Fedotov is recovered, where the
equation
\begin{equation}
a(0) = 1\, \label{102}
\end{equation}
and
\begin{equation}
\frac{g_{1}^{2}}{\omega_{0}\Omega}
\tanh\biggl(\frac{\beta_{c}\,\Omega}{4}\biggr) = 1\,, \label{103}
\end{equation}
give the inverse of the critical temperature, $\beta_{c}$. It is
given by
\begin{equation}
\beta_{c} =
\frac{4}{\Omega}\tanh^{-1}\biggl(\frac{\omega_{0}\Omega}{g_{1}^{2}}\biggr)\,
. \label{104}
\end{equation}
In this case, also there is a quantum phase transition, i.e., a
zero temperature phase transition when
$g_{1}=(\omega_{0}\,\Omega)^{\frac{1}{\,2}}$. It is interesting to
point out that there are two different ways to analyze the phase
transition. The first one is to follow the non-analytic behavior
of the thermodynamic quantities as a function of temperature. A
different way is to follow the non-analytic behavior of the
thermodynamic quantities as a function of the coupling constant
strength. Working in this second approach, we may expect that for
large coupling constant $g_{1}$ there is a superradiant phase. The
spectrum of the collective Bose excitations in this case is
\begin{equation}
E_{1}=0\, , \label{108}
\end{equation}
and
\begin{equation}
E_{2}=\Omega+\omega_{0}\, . \label{109}
\end{equation}
Now we will show that it is possible to have a condensate with
superradiance in a system of $N$ qubits coupled with one mode of a
Bose field where only virtual processes contribute.
In the pure counter-rotating wave case, i.e., $g_{1}=0$ and
$g_{2}\neq 0$, the inverse of the critical temperature,
$\beta_{c}$ is given by
\begin{equation}
\beta_{c} =
\frac{4}{\Omega}\tanh^{-1}\biggl(\frac{\omega_{0}\Omega}{g_{2}^{2}}\biggr)\,
. \label{112}
\end{equation}
and the spectrum of the colective Bose excitations given by
\begin{equation}
E_{1}=0\, , \label{110}
\end{equation}
and
\begin{equation}
E_{2}=|\,\Omega-\omega_{0}|\, . \label{111}
\end{equation}

A comment is in order concerning the Bose excitations spectrum. In
the both cases: using or not the rotating-wave approximation,
there is a phase transition. In the case of the rotating-wave
approximation $g_{1}\neq 0$ and $g_{2}=0$, there is a Goldstone
mode $(E=0)$. In the pure counter-rotating wave case $g_{1}=0$ and
$g_{2}\neq 0$, also there is a Goldstone (gapless) mode. The
existence of Goldstone modes and the energy of the other mode was
presented for both above mentioned cases. The spectrum in the
general case is given by the Goldstone mode and also by a non-zero
energy mode given by Eq. (\ref{107}). It is interesting to stress
that we obtained a critical behavior in both situations
$(g_{1}\neq 0$, $g_{2}=0$ and $g_{1}=0$, $g_{2}\neq 0)$, where the
condensate has Goldstone (gapless) modes, with a superradiant
state. Therefore we show that it is possible to have a condensate
with superradiance in a system of $N$ qubits coupled with one mode
of a Bose field where only virtual processes contribute.

We would like to point out to the reader that there is an analogy
between the generalized Dicke model and the Landau-Ginzburg model
in the theory of phase transition. To describe the second-order
phase transition Landau introduced the concept of order parameter.
This concept enabled Landau to write the free energy of any system
in the form of an expansion in powers of the order parameter. This
expansion characterizes order in the system as a whole. A
generalization of this model accounts for the local space
variation, i.e., to take into account possible local fluctuation
of the order parameter. In the Landau theory we ignore
fluctuations as the same way that in the rotating-wave
approximation in the Dicke model. Going along this line we can
make a parallel between the Landau-Ginzburg model and the
generalized Dicke model. In concluding we remark that the model
with pure virtual processes is enable to generate a condensate.

\section{Conclusions}

\quad $\,\,$In the present paper we are first discussing
Hamiltonians describing a quantized bosonic field interacting with
two-level atoms. Without assuming the rotating-wave-approximation,
with the coupling constants $g_{1}$ and $g_{2}$ for rotating and
counter-rotating terms respectively, we define the generalized
fermion Dicke model. Changing the atomic pseudo-spin operators of
the model by a linear combination of Fermi operators we define the
fermion generalized Dicke model. Studying the case where identical
two-level atoms act as a thermal reservoir $(N \rightarrow
\infty)$, we investigate the thermodynamic of the generalized
Dicke model using the path integral approach with functional
integration method. We are considering the question of how does
the counter-rotating terms of the interaction Hamiltonian
contribute in the formation of the condensate with a superradiant
phase transition in the model.

We study the nonanalytic behavior of thermodynamic quantities of
the generalized model for both situations, evaluating the critical
transition temperature and presenting the spectrum of the
colective bosonic excitations, for the case $g_{1}\neq 0$ and
$g_{2}=0$, $g_{1}=0$ and $g_{2}\neq 0$ and also in the general
case. Our result show that it is possible to have a condensate
with superradiance in a system of $N$ qubits coupled with one mode
of a bosonic field where only virtual processes contribute. It is
important to realize that the energy of the non-Goldstone mode in
Eq.(\ref{109}) is always larger than the energy of the
non-Goldstone mode of Eq. (\ref{111}), i.e., in the system where
the condensate appears due to the virtual processes. Our
conclusion from the above results is that both processes, real and
virtual ones give different contributions to generate the
condensate.

An important question is the way of practical realization of the
generalized Dicke model in the laboratory. As was stressed by
Dimer et al \cite{dimer} it remains as a challenge to provide a
physical system where the counter-rotating terms are dominant.
Those authors proposed that in cavities with the $N$ qubits, only
one mode of quantized field and classical fields (lasers), it is
possible to obtain a physical system that corresponds to the
generalized Dicke model. Also, has been discussed in the
literature the possibility of control the relative importance of
the counter-rotating terms in the Jaynes-Cummings model in the
laboratory using a ion trap \cite{cirac}. Another mechanism to
explore the importance of virtual processes was proposed by Ford
\cite{foc1} and Ford and Svaiter \cite{foc2} \cite{foc3}, where
the possibility of amplification of the vacuum fluctuations was
discussed. These authors studied the renormalized vacuum
fluctuations associated with a scalar and electromagnetic field
near the focus of a parabolic mirror. Using the geometric optics
approximation these authors found that the mirror geometry can
produce large vacuum fluctuations near the focus.

It is not our proposal to discuss the practical realization of the
generalized Dicke model in the laboratory, but only analyze the
thermodynamic of the model using the path integral approach with
functional method. An evidence in favor of our results is an
experiment where it is possible to control the importance of the
counter-rotating terms in the generalized model in such a way that
an ideal $g_{1}\approx 0$ situation is achieved. Experimental
observation of the superradiant phase in this situation will
improve our understanding of this phenomenon, when the atoms
radiates spontaneously, at a radiation rate much higher than would
be expected from an ensemble of independently radiating atoms.

 There are many different continuations for
this paper. The first one is to investigate the model introduced
by DiVicenzo, defined by Eq. (\ref{31}), at finite temperature,
using also functional integral methods. Also the model defined by
Eq. (\ref{a28}), with a coupling between the $N$ qubits and the
single quantized mode of a bosonic field, which characterize a
intensity-dependent coupling  can  be discussed using the
conventional technique of the path integral and functional
integration method. A similar calculation presented in this paper
could be carried out for this model, although the calculations
would be somewhat more laborious that the presented in the paper.
The generalization of this model with the introduction of the
counter-rotating terms deserves further investigations.

An outstanding question is the presence of the quantum phase
transition and also the chaotic behavior for a finite $N$, where a
crossover between Poisson and Wigner-Dyson behavior in the energy
level-spacing statistics appears. The question that has been
discussed in the literature is the distribution function of
spacing of adjacent eigenenergies in a large number of systems.
For systems where the classical dynamics is integrable, a
Poissonian nearest neighbor spacing distribution is expected,
i.e., the spacing rule for random levels. Some systems, the
eigenenergies do not follow the Poisson law, but behave as the
eigenvalues of a random matrix taken from a suitable ensemble. The
set of all real random matrices with matrix elements obeying some
distribution function defines the Gaussian orthogonal ensemble.
Other ensembles are the Gaussian unitary and the Gaussian
sympletic ensemble respectively. The chaotic behavior for a finite
$N$ in the above discussed model is under investigation by the
authors.

\section{Acknowlegements}

We would like to thanks Alexis Hern\'andez for useful discussions.
This paper was supported by Conselho Nacional de Desenvolvimento
Cientifico e Tecnol{\'o}gico do Brazil (CNPq).

\end{document}